\documentclass[fleqn,10pt]{wlscirep}
\usepackage[utf8]{inputenc}
\usepackage[T1]{fontenc}
\usepackage[utf8]{inputenc}
\usepackage[T1]{fontenc}
\usepackage{float}
\usepackage{amsmath,amssymb,amsfonts}
\usepackage{array,multirow,graphicx}
\usepackage{float}
\usepackage{algorithmic}
\usepackage{graphicx}
\usepackage{textcomp}
\usepackage{xcolor}
\usepackage{hyperref}
\usepackage{cleveref}
\usepackage{footmisc}
\usepackage{xcolor,colortbl}

\definecolor{Gray}{gray}{0.85}
\definecolor{LightCyan}{rgb}{0.88,1,1}

\newcolumntype{a}{>{\columncolor{Gray}}c}
\newcolumntype{b}{>{\columncolor{white}}c}

\title{PediCXR: An open, large-scale chest radiograph dataset for interpretation of common thoracic diseases in children}

\author[1,3,4,*]{Hieu H. Pham}
\author[2]{Ngoc H. Nguyen}
\author[1]{Thanh T. Tran}
\author[5]{Tuan N.M. Nguyen}
\author[1]{Ha Q. Nguyen}
\affil[1]{Smart Health Center, VinBigData JSC,  Hanoi, Vietnam}
\affil[2]{Phu Tho Department of Health, Phu Tho, Vietnam}
\affil[3]{College of Engineering \& Computer Science, VinUniversity,  Hanoi, Vietnam}
\affil[4]{VinUni-Illinois Smart Health Center, Hanoi, Vietnam}
\affil[5]{Training and Direction of Healthcare Activities Center, Phu Tho General Hospital, Phu Tho, Vietnam}
\affil[*]{These authors contributed equally:  Hieu H. Pham and Ngoc H. Nguyen}
\affil[*]{Corresponding author: Hieu H. Pham (hieu.ph@vinuni.edu.vn)} 

\begin{abstract}
Computer-aided diagnosis systems in adult chest radiography (CXR) have recently achieved great success thanks to the availability of large-scale, annotated datasets and the advent of high-performance supervised learning algorithms. However, the development of diagnostic models for detecting and diagnosing pediatric diseases in CXR scans is undertaken due to the lack of high-quality physician-annotated datasets. To overcome this challenge, we introduce and release PediCXR, a new pediatric CXR dataset of 9,125 studies retrospectively collected from a major pediatric hospital in Vietnam between 2020 and 2021. Each scan was manually annotated by a pediatric radiologist with more than ten years
of experience. The dataset was labeled for the presence of 36 critical findings and 15 diseases. In particular, each abnormal finding was identified via a rectangle bounding box on the image. To the best of our knowledge, this is the first and largest pediatric CXR dataset containing lesion-level annotations and image-level labels for the detection of multiple findings and diseases. For algorithm development, the dataset was divided into a training set of 7,728 and a test set of 1,397. To encourage new advances in pediatric CXR interpretation using data-driven approaches, we provide a detailed description of the PediCXR data sample and make the dataset publicly available on \href{https://physionet.org/content/vindr-pcxr/1.0.0/}{https://physionet.org/content/pedicxr/1.0.0/}.
\end{abstract}

\begin{document}

\flushbottom
\maketitle

\thispagestyle{empty}

\section*{Background \& Summary} \label{sec:introduction}

Common thoracic diseases cause several hundred thousand deaths every year among children under five years old~\cite{GBD2015,unicef2006}. The chest radiograph or CXR is the first-line and most commonly performed imaging examination in the assessment of the pediatric patient~\cite{hart2019pediatric}. Interpreting CXR scans on pediatric patients can be for a number of indications or critical findings, in particular for common thoracic diseases in children such as Pneumonia, Bronchitis  and Cardiovascular diseases (CVDs). Depending on the patients' age, the difficulty of the examination will vary, often requiring a specialist in pediatric diagnostic imaging with an in-depth knowledge of radiological signs of different lung conditions~\cite{pcxr}. Additionally, the inter-observer agreement and intra-observer agreement in the pediatric CXR interpretation were low~\cite{du2002observer}. This opens room for the development of data-driven approaches and computational tools to assist pediatricians in the diagnosis of common thoracic diseases and to reduce their workload. 

Computer-aided diagnosis (CAD) systems for identification of lung abnormality in adult CXRs have recently achieved great success thanks to the availability of large labeled datasets~\cite{wang2017chestx,bustos2019padchest,irvin2019chexpert,johnson2019mimic,nguyen2020vindr}. Many large-scale CXR datasets of adult patients such as Montgomery County chest X-ray (MC)~\cite{jaeger2014two}, Shenzhen chest X-ray~\cite{jaeger2014two}, ChestX-ray8~\cite{wang2017chestx}, COVIDGR~\cite{tabik2020covidgr}, ChestX-ray14~\cite{wang2017chestx}, Padchest~\cite{bustos2019padchest}, CheXpert~\cite{irvin2019chexpert}, MIMIC-CXR~\cite{johnson2019mimic} and VinDr-CXR~\cite{nguyen2020vindr} have been established and released in recent years. These datasets boosted new advances in exploring new machine learning-based approaches in the interpretation of CXR in adults~\cite{rajpurkar2017chexnet,rajpurkar2018deep,irvin2019chexpert,majkowska2020chest,rajpurkar2020chexpedition,tang2020automated,pham2020interpreting}. Unfortunately, the creation of pediatric CXR datasets is still unexploited, and the number of benchmark pediatric CXR datasets is limited. This becomes the main obstacle in developing and transferring new machine learning-based CAD systems for pediatric CXR in clinical practice.

In an effort to provide a large-scale pediatric CXR dataset with high-quality annotations for the research community, we have built the PediCXR dataset in DICOM format. The dataset consists of 9,125 posteroanterior (PA) view CXR scans in patients younger than 10 years that were retrospectively collected from three major hospitals in Vietnam from 2020 to 2021. In particular, all CXR scans come with both the localization of critical findings and the classification of common thoracic diseases. These images were annotated by a group of three radiologists with at least 10 years of experience for the presence of 36 critical findings (\textit{local labels}) and 15 diagnoses (\textit{global labels}). Here, the local labels should be annotated with rectangle bounding boxes that localize the findings, while the global labels reflect the diagnostic impression of the radiologist at the image-level. For algorithm development, we randomly divided the dataset into two parts: the training set of 7,728 scans (84.7\%) and the test set of 1,397 scans (15.3\%). To the best of our knowledge, the released PediCXR is currently the largest public pediatric CXR dataset with radiologist-generated annotations in both training and test sets. Table~\ref{existing-data} below shows an overview of existing public datasets for CXR interpretation in pediatric patients, compared with the PediCXR. Compared to the previous works, the PediCXR dataset shows two main advantages. First, the dataset is labeled for multiple findings and diseases. Meanwhile, most pediatric CXR datasets have focused on a single disease such as pneumonia~\cite{KERMANY20181122} or pneumothorax~\cite{Chen2020}. Second, the dataset provides bounding box annotations at lesion level, which is useful for developing explainable artificial intelligent models~\cite{gordon2019explainable} for the CXR interpretation in children. We believe the introduction of the PediCXR provides a suitable imaging source for investigating the ability of supervised machine learning models in identifying common lung diseases in pediatric patients.

\begin{table}[ht]
\textsf{\scriptsize{
\caption{\textsf{An overview of existing public datasets for CXR interpretation in pediatric patients.}}
\label{existing-data}
\setlength{\tabcolsep}{3pt}
\begin{tabular}{ p{90pt}|p{70pt}|p{70pt}|p{70pt}|p{70pt}|p{90pt}}
\hline
 \textbf{Dataset} & \hspace*{0.5cm} \textbf{Release year} & 
\hspace*{0.5cm} \# \textbf{findings} &  \hspace*{0.5cm} \# \textbf{samples} &  \hspace*{0.2cm} \textbf{Image-level labels} &  \hspace*{0.85cm} \textbf{Local labels}\\
\hline
Kermany \textit{et al.}~\cite{KERMANY20181122}& \hspace*{1cm} 2018 & \hspace*{1cm} 2 & \hspace*{0.7cm} 5,856  & \hspace*{0.7cm} Available & \hspace*{1cm} Not available\\
 Chen \textit{et al.}~\cite{Chen2020} &  \hspace*{1cm} 2020 &  \hspace*{1cm} 5 &  \hspace*{0.7cm} 2,668 &  \hspace*{0.7cm} Available &  \hspace*{1cm} Not available\\
\hline
 \textbf{PediCXR (ours)}   &  \hspace*{1cm} \textbf{2021} &  \hspace*{1cm} \textbf{52} &  \hspace*{0.7cm} \textbf{9,125}  &  \hspace*{0.7cm} \textbf{Available} &  \hspace*{1cm} \textbf{Available} \\
\hline
\end{tabular}}}
\end{table}

 \section*{Methods}
 
\subsection*{Data collection}
\label{sec:collect}
Data collection was conducted at the Phu Tho Obstetric \& Pediatric Hospital (PTOPH) between 2020 -- 2021. The ethical clearance of this study was approved by the Institutional Review Boards (IRBs) of the PTOPH. The need for obtaining informed patient consent was waived because this retrospective study did not impact clinical care or workflow at these two hospitals, and all patient-identifiable information in the data has been removed. We retrospectively collected more than 10,000 CXRs in DICOM format from a local picture archiving and communication system (PACS) at PTOPH. The imaging dataset was then transferred and analyzed at Smart Health Center, VinBigData JSC.

\subsection*{Overview of approach}
\label{sec:approach}
\begin{figure}[H]
\centerline{\includegraphics[width=12cm,height=5.5cm]{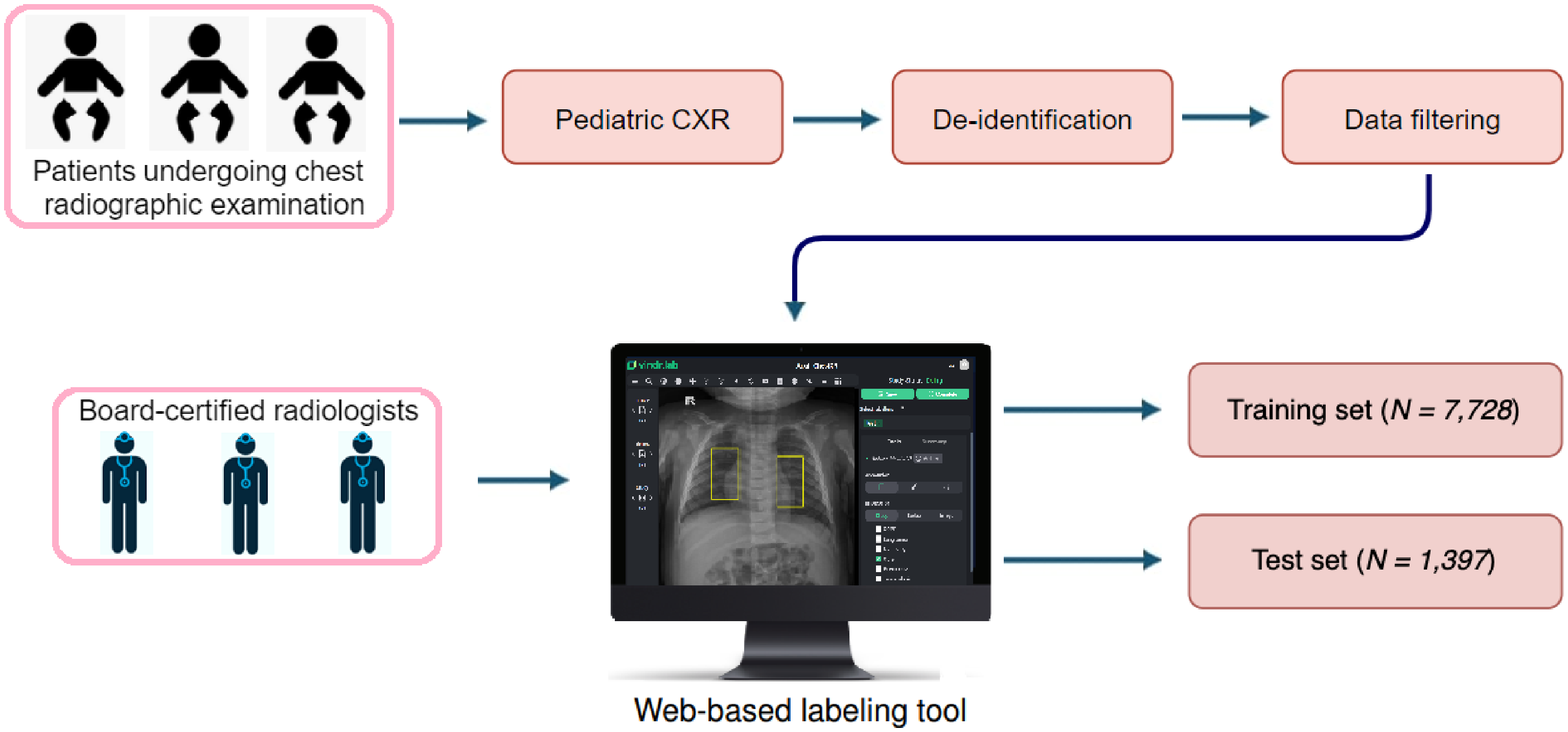}}
\caption{{\textsf{\small{Construction of the PediCXR dataset. First, raw pediatric scans in DICOM format were collected retrospectively from the hospital's PACS at PTOPH. These images were de-identified to protect patient's privacy. Then, invalid files (including adult CXR images, images of other modalities or other body parts, images with low quality, or incorrect orientation) were manually filtered out. After that, a web-based DICOM labeling tool called VinDr Lab was developed to remotely annotate DICOM data. Finally, the annotated dataset was then divided into a training set (\textit{N = 7,728}) and a test set (\textit{N = 1,397}) for algorithm development.}}}}
\label{fig:data_colection}
\end{figure}

The building of the PediCXR dataset is illustrated in Figure~\ref{fig:data_colection}. In particular, the collection and normalization of the dataset were divided into four main steps: (1) data collection, (2) data de-identification, (3) data filtering, and (4) data labeling. We describe each step in detail as below.

\subsection*{Data de-identification}
\label{sec:dataset}
In this study, we follow the HIPAA Privacy Rule~\cite{assistance2003summary} to protect individually identifiable health information from the DICOM images. To this end, we removed or replaced with random values all personally identifiable information associated with the images via a two-stage de-identification process. At the first stage, a Python script was used to remove all DICOM tags of protected health information (PHI)~\cite{isola2019protected} such as patient's name, patient's date of birth, patient ID, or acquisition time and date, etc. For the purpose of loading and processing DICOM files, we only retained a limited number of DICOM attributes that are necessary, as indicated in Table~\ref{dicom_tags_retained} (Supplementary materials). In the second stage, we manually removed all textual information appearing on the image data, i.e., pixel annotations that could include patient's identifiable information.

\begin{table*}
\centering
\textsf{\scriptsize{
\caption{\textsf{The list of DICOM tags that were retained for loading and processing raw images. All other tags were removed for protecting patient privacy. Details about all these tags can be found from DICOM Standard Browser at \href{https://dicom.innolitics.com/ciods}{https://dicom.innolitics.com/ciods}.}}
\label{dicom_tags_retained}
\setlength{\tabcolsep}{3pt}
\begin{tabular}{p{130pt}|p{150pt}|p{200pt}}
\\
\hline
\hspace*{1cm} \textbf{DICOM Tag} & 
\hspace*{1cm}\textbf{Attribute Name} & \textbf{Description} \\
\hline
& & \\
\hspace*{1cm} (0010, 0040) & \hspace*{1cm} Patient's Sex & Sex of the named patient.\\
& & \\
\hspace*{1cm} (0010, 1010) & \hspace*{1cm} Patient's Age & Age of the patient.\\
& & \\
\hspace*{1cm} (0010, 1020) & \hspace*{1cm} Patient's Size & Length or size of the patient, in meters.\\
& \\
\hspace*{1cm} (0010, 1030) & \hspace*{1cm} Patient's Weight & Weight of the patient, in kilograms.\\
& \\
\hspace*{1cm} (0028, 0010) & \hspace*{1cm} Rows & Number of rows in the image.\\
& \\
\hspace*{1cm} (0028, 0011) & \hspace*{1cm} Columns & Number of columns in the image.\\
& \\
\hspace*{1cm} (0028, 0030) & \hspace*{1cm} Pixel Spacing & Physical distance in the patient between the center of each pixel, specified by a numeric pair - adjacent row spacing (delimiter) adjacent column spacing in mm. \\
& \\
\hspace*{1cm} (0028, 0034) & \hspace*{1cm} Pixel Aspect Ratio & Ratio of the vertical size and horizontal size of the pixels in the image specified by a pair of integer values where the first value is the vertical pixel size, and the second value is the horizontal pixel size.\\
& \\
\hspace*{1cm} (0028, 0100) & \hspace*{1cm} Bits Allocated & Number of bits allocated for each pixel sample. Each sample shall have the same number of bits allocated. \\
& \\
\hspace*{1cm} (0028, 0101) & \hspace*{1cm} Bits Stored & Number of bits stored for each pixel sample. Each sample shall have the same number of bits stored.\\
& \\
\hspace*{1cm} (0028, 0102) & \hspace*{1cm} High Bit & Most significant bit for pixel sample data. Each sample shall have the same high bit. \\
& \\
\hspace*{1cm} (0028, 0103) & \hspace*{1cm} Pixel Representation & Data representation of the pixel samples. Each sample shall have the same pixel representation.\\
& \\
\hspace*{1cm} (0028, 0106) & \hspace*{1cm} Smallest Image Pixel Value & The minimum actual pixel value encountered in this image.\\
& \\
\hspace*{1cm} (0028, 0107) & \hspace*{1cm} Largest Image Pixel Value & The maximum actual pixel value encountered in this image.\\
& \\
\hspace*{1cm} (0028, 1050) & \hspace*{1cm} Window Center & Window center for display.\\
& \\ 
\hspace*{1cm} (0028, 1051) & \hspace*{1cm} Window Width & Window width for display.\\
& \\
\hspace*{1cm} (0028, 1052) & \hspace*{1cm} Rescale Intercept & The value b in relationship between stored values (SV) and the output units specified in Rescale Type (0028,1054). Each output unit is equal to m*SV + b.\\
& \\
\hspace*{1cm} (0028, 1053) & \hspace*{1cm} Rescale Slope & Value of m in the equation specified by Rescale Intercept (0028,1052).\\
& \\
\hspace*{1cm} (7FE0, 0010) & \hspace*{1cm} Pixel Data & A data stream of the pixel samples that comprise the image.\\
& \\
\hspace*{1cm} (0028, 0004) & \hspace*{1cm} Photometric Interpretation & Specifies the intended interpretation of the pixel data.\\
& \\
\hspace*{1cm} (0028, 2110) & \hspace*{1cm} Lossy Image Compression & Specifies whether an image has undergone lossy compression (at a point in its lifetime).\\
&\\
\hspace*{1cm} (0028, 2114) & \hspace*{1cm} Lossy Image Compression Method & A label for the lossy compression method(s) that have been applied to this image.\\
&\\
\hspace*{1cm} (0028, 2112) & \hspace*{1cm} Image Compression Ratio & Describes the approximate lossy compression ratio(s) that have been applied to this image.\\
&\\
\hspace*{1cm} (0028, 0002) & \hspace*{1cm} Samples per Pixel & Number of samples (planes) in this image. \\
&\\
\hspace*{1cm} (0028, 0008) & \hspace*{1cm} Number of Frames & Number of frames in a multi-frame image.\\
&\\
\hline
\end{tabular}
}}
\end{table*}

\subsection*{Data filtering}
\label{sec:dataset}
The collected raw data included a significant amount of outliers including CXRs of adult patients, body parts other than chest (abdominal, spine, and others),  low-quality images, or lateral CXRs. To filter a large number of CXR scans, we trained a lightweight convolutional neural network (CNN)~\cite{pham2021dicom} to remove all outliers automatically. Next, a manual verification was performed to ensure all outliers had been fully removed.

\subsection*{Data labeling}
\begin{figure}[ht]
\centerline{\includegraphics[width=\linewidth]{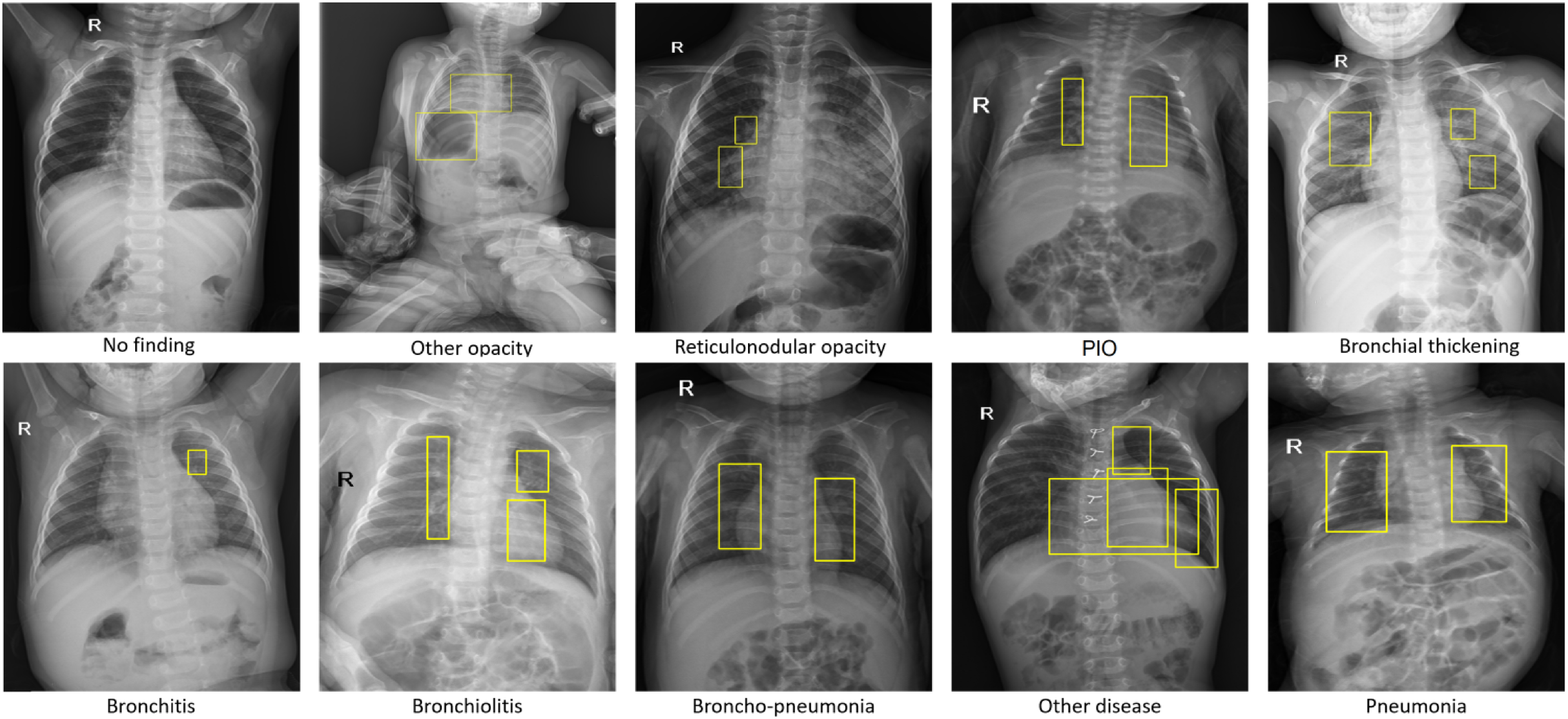}}
\caption{{\textsf{\small{Several examples of pediatric CXR images with radiologist's annotations. Local labels marked by radiologists are plotted on the original images for visualization purposes. These annotations show abnormal findings from the scans. The global labels, that classify images into diseases, are in bold and listed at the bottom of each example.}}}}
\label{representive_cases}
\end{figure}

The PediCXR dataset was labeled for a total of 36 findings and 15 diagnoses. These labels were divided into two categories: local labels (\#1 -- \#36) and global labels (\#37 -- \#52). The local labels should be marked with bounding boxes that localize the findings, while the global labels should reflect the diagnostic impression of the radiologist. This list of labels was suggested by a committee of the most experienced pediatric radiologists. To select these labels, the committee took into account two key factors. First, findings and diseases are prevalent. Second, they can be differentiated on pediatric chest X-ray scans. Figure~\ref{representive_cases} illustrates several samples with both local and global labels annotated by our radiologists.

To facilitate the labeling process, we designed and built a web-based framework called VinDr Lab~\cite{nguyen2021vindr} that allows a team of experienced radiologists remotely annotate the data. Specifically, this is a web-based labeling tool that was developed to store, manage, and remotely annotate DICOM data. The radiologists were oriented to locate the abnormal findings from the DICOM viewer and draw the bounding boxes. All the annotators have been well-trained to ensure that the annotations are consistently annotated. In addition, all the radiologists participating in the labeling process were certified in diagnostic radiology and received healthcare professional certificates. In total, three pediatric radiologists with at least 15 years
of experience were involved in the annotation process. Each sample in the training set was assigned to one radiologist for annotation. Additionally, all of the participating radiologists were blinded to relevant clinical information. A set of 9,125 pediatric CXRs were randomly annotated from the filtered data, of which 7,728 scans serve as the training set, and the remaining 1,397 studies form the test set. Note the 9,125 studies correspond to 9,125 patients, and each study has a single CXR scan.

\begin{table}[H]
\centering
\textsf{\scriptsize{
\caption{\textsf{Dataset characteristics of PediCXR.}}
\label{data-characs}
\setlength{\tabcolsep}{3pt}
\begin{tabular}{p{15pt} | p{180pt}|p{100pt}|p{120pt}}
\hline
\hline
&  \textbf{Characteristics} & 
\textbf{Training set} & 
\textbf{Test set}  \\
\hline
  \parbox[t]{2mm}{\multirow{10}{*}{\rotatebox[origin=c]{90}{\textbf{Collection statistics}}}} & &  \\
  &   Years &  2020 to 2021 &  2020 to 2021\\
&  Number of scans & 7,728 & 1,397 \\
&   Number of human annotators per scan  &  1 &  1  \\
&  Image size (pixel$\times$pixel, median) &  1,643 $\times$ 1,349 &  1,638 $\times$ 1,343 \\
&   Age (years, median)*  &  1.71 &  1.69  \\
&   Male (\%)* &  57.63  & 59.14    \\
&   Female (\%)* &  42.37  &  40.86  \\
&  Data size (GB) &   30.9 &   5.7 \\
& & &  \\
\hline
\parbox[t]{3mm}{\multirow{22.5}{*}{\rotatebox[origin=c]{90}{\textbf{Local labels}}}}  & &   \\
&  1. Boot-shaped heart (\%)  &  35 (0.45\%)  &    6 (0.43\%) \\
&                       2. Peribronchovascular interstitial opacity or PIO (\%)         &  1,358 (17.57\%)                          & 248 (17.75\%)  \\
&  3. Reticulonodular opacity (\%)        &  509 (6.59\%)  &   90 (6.44\%) \\
&                       4. Bronchial thickening (\%)       & 562 (7.27\%)                          & 116 (8.30\%) \\
&  5. Enlarged PA (\%)   &  61 (0.79\%)      &  11 (0.79\%) \\
&                       6. Cardiomegaly (\%)       &  161 (2.08\%)                         & 29 (2.08\%) \\
&  7. Other opacity (\%)               &  148 (1.92\%)      &   27 (1.93\%) \\
&                       8. Intrathoracic digestive structure (\%)           &  2 (0.03\%)                          & 0 (0.00\%) \\
&  9. Diffuse aveolar opacity (\%)         &  119 (1.54\%)     &  21 (1.50\%) \\
&                       10. Other lesion (\%)       &  65 (0.84\%)    &  11 (0.79\%) \\
&  11. Consolidation  (\%)       &  176 (2.28\%)    &  35 (2.51\%) \\
&                       12. Mediastinal shift (\%)          &  5 (0.06\%)     &  0 (0.00\%) \\
&  13. Anterior mediastinal mass (\%)          &   5 (0.06\%)     &  1 (0.07\%) \\
&                       14. Other nodule/mass (\%)  &  10 (0.13\%)     &  2 (0.14\%) \\
&  15. Dextro cardia (\%)  &  16 (0.21\%)     &  3 (0.21\%) \\
&                       16. Aortic enlargement (\%) &  2 (0.03\%)   &  0 (0.00\%) \\
&  17. Pleural effusion (\%) &   14 (0.18\%)   &  3 (0.21\%) \\
&                       18. Stomach on the right side (\%) &  5 (0.06\%)   &  1 (0.07\%) \\
&  19. Atelectasis (\%) &   23 (0.30\%)   &  3 (0.21)\%) \\
&                       20. Calcification (\%)       &  1 (0.01\%)     &  0 (0.00\%) \\
&  21. Interstitial lung disease - ILD (\%)       &  14 (0.18\%)     &  2 (0.14\%) \\
&                       22. Lung hyperinflation (\%)       &  108 (1.40\%)     &  21 (1.50\%) \\
&  23. Egg on string sign (\%)       &  12 (0.16\%)     &  2 (0.14\%) \\
&                       24. Pulmonary fibrosis (\%)       &  1 (0.01\%)     &  0 (0.00\%) \\
&  25. Infiltration (\%)       &  11 (0.14\%)     &  2 (0.14\%) \\
&                       26. Lung cavity (\%)       &  5 (0.06\%)     &  1 (0.07\%) \\
&  27. Pneumothorax (\%)       &  4 (0.05\%)     &  0 (0.00\%) \\
&                       28. Edema (\%)       &  1 (0.01\%)     &  0 (0.00\%) \\
&  29. Pleural thickening (\%)       &  2 (0.03\%)     &  0 (0.00\%) \\
&                       30. Clavicle fracture (\%)       &  5 (0.06\%)     &  1 (0.07\%) \\
&  31. Chest wall mass (\%)       &  3 (0.04\%)     &  0 (0.00\%) \\
&                       32. Lung cyst (\%)       &  8 (0.10\%)     &  2 (0.14\%) \\
&  33. Emphysema (\%)       &  1 (0.01\%)     &  0 (0.00\%) \\
&                       34. Bronchectasis (\%)       &  3 (0.04\%)   &   0 (0.00\%) \\
&  35. Expanded edges of the anterior ribs (\%)       &  2 (0.03\%)     &  0 (0.00\%) \\
&                       36. Paraveterbral mass (\%)       &  2 (0.03\%)     &  0 (0.00\%) \\
& & & 
\\
\hline 
\parbox[t]{2mm}{\multirow{8}{*}{\rotatebox[origin=c]{90}{\textbf{Global labels}}}} & &  \\
&  37. No finding (\%)         &  5,143 (66.55\%)     &  907 (64.92\%) \\
& 38. Bronchitis (\%)                                &  842 (10.90\%)                         &  174 (12.46\%) \\
&  40. Brocho-pneumonia (\%)       &  545 (7.05\%)    &  84 (6.01\%)  \\
&  41. Other diseases (\%)                          &   412 (5.33\%)                      &  77 (5.51\%) \\
&  42. Bronchiolitis (\%)               &  497 (6.43\%)      &   90 (6.44\%) \\
&  43. Situs inversus (\%)                              & 11 (0.14\%)                       &  2 (0.14\%)  \\
&  44. Pneumonia (\%)        &   392 (5.07\%)   &  89 (6.37\%) \\
& 45. Pleuro-pneumonia (\%)                                &  6 (0.08\%)                         &  0 (0.00\%) \\
&  46. Diagphramatic hernia (\%)       &  3 (0.04\%)    &  0 (0.00\%)  \\
&  47. Tuberculosis (\%)                          &   14 (0.18\%)                      &  1 (0.07\%) \\
&  48. Congenital emphysema (\%)               &  2 (0.03\%)      &   0 (0.00\%) \\
&  49. CPAM (\%)                              & 5 (0.06\%)                       &  1 (0.07\%)  \\
&  50. Hyaline membrane disease (\%)               &  19 (0.25\%)      &   3 (0.21\%) \\
&  51. Mediastinal tumor (\%)                              & 8 (0.10\%)                       &  1 (0.07\%)  \\
&  52. Lung tumor (\%)               &  5 (0.06\%)      &   0 (0.00\%) \\
\hline
\hline
\end{tabular}
}}
\end{table}

\begin{figure}[H]
\centerline{\includegraphics[width=17cm,height=7cm]{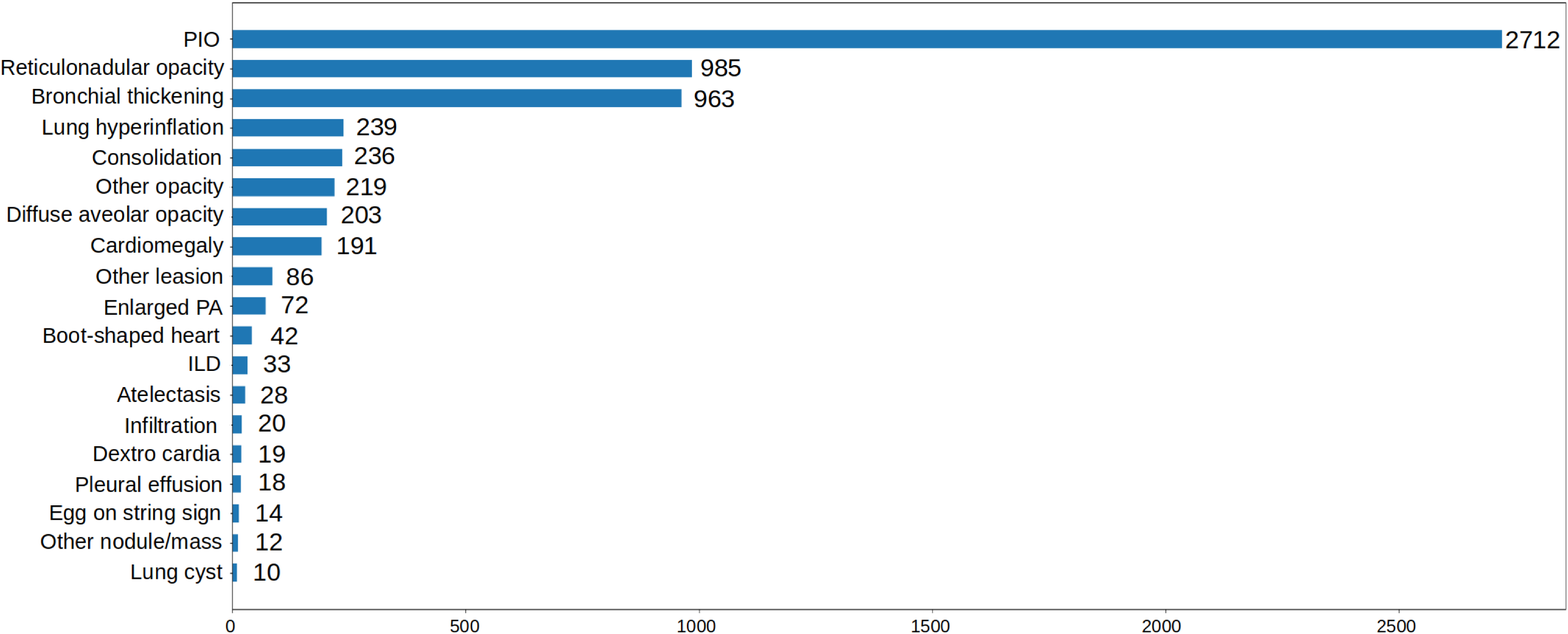}}
\caption{{\textsf{\small{Distribution of abnormal findings on the training set of PediCXR. Rare findings (less than 10 examples) are not included.}}}}
\label{label-distribution-1}
\end{figure}

\begin{figure}[H]
\centerline{\includegraphics[width=17cm,height=6cm]{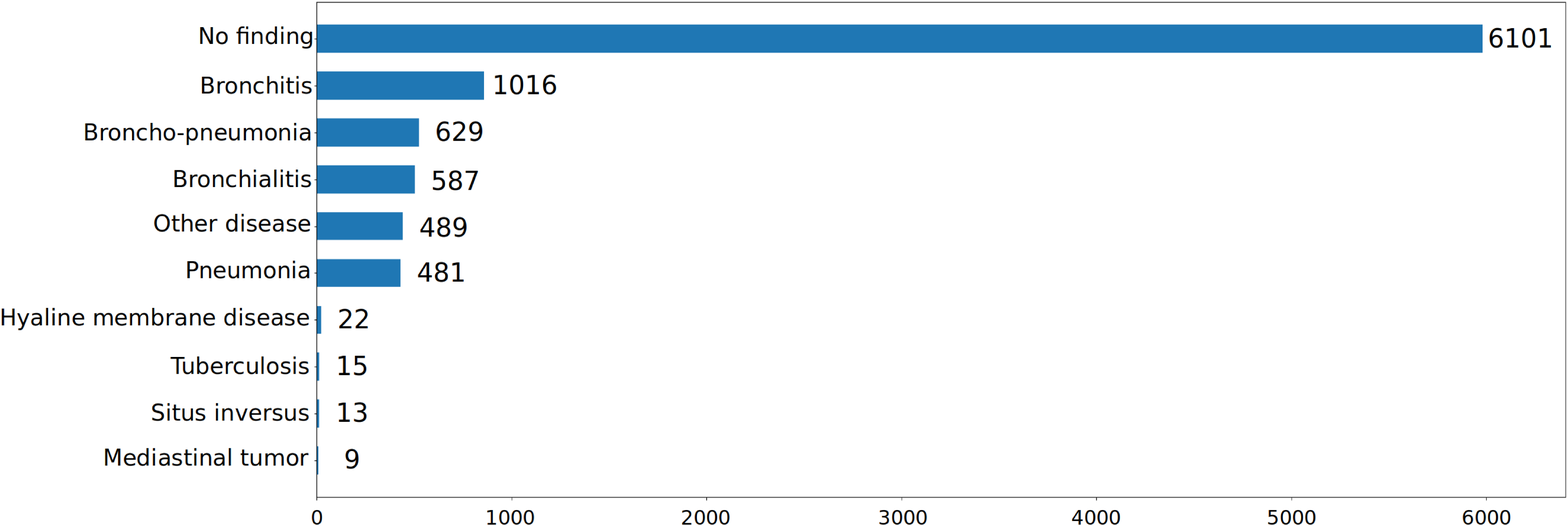}}
\caption{{\textsf{\small{Distribution of pathologies on the training set of PediCXR. Rare diseases (less than 10 examples) are not included.}}}}
\label{label-distribution-2}
\end{figure}

Once the labeling was completed, the annotations of all pediatric CXRs were exported in JavaScript Object Notation (JSON) format. We developed a Python script to parse JSON files and organized the annotations in the form of a single comma-separated values (CSV) file. Each CSV file contains labels, bounding box coordinates, and their corresponding image identifiers (IDs). The data characteristics, including patient demographic and the prevalence of each finding or disease, are summarized in Table~\ref{data-characs}. The distributions of abnormal findings and pathologies in the training set are drawn in Figure~\ref{label-distribution-1} and Figure~\ref{label-distribution-2}, respectively. 

\section*{Data Records}
\label{sec:data_records}

The PediCXR dataset will be made available for public download on PhysioNet~\cite{data_physi}. We offer complete imaging data as well as ground truth labels for both the training and test datasets. The pediatric scans were split into two folders: one for training and one for testing, named as ``\verb|train|'' and ``\verb|test|'', respectively. Since each study has only one instance and each patient has maximum one study, therefore, the value of the SOP Instance UID provided by the DICOM tag $(0008,0018)$ was encoded into a unique, anonymous identifier for each image. To this end, we used the Python \verb|hashlib| module (see \hyperref[sec:code]{Code Availability}) to encode the SOP Instance UIDs into image IDs. The radiologists' local annotations of the training set were provided in a CSV file called \verb|annotations_train.csv|. Each row of the CSV file represents a bounding box annotation with the following attributes: image ID (\verb|image_id|), radiologist ID (\verb|rad_id|), label's name (\verb|class_name|), bounding box coordinates (\verb|x_min|, \verb|y_min|, \verb|x_max|, \verb|y_max|), and label class ID (\verb|class_id|). The coordinates of the box's upper-left corner are (\verb|x_min|, \verb|y_min|), and the coordinates of the box's lower right corner are (\verb|x_max|, \verb|y_max|). Meanwhile, the image-level labels of the training set were stored in a different CSV file called \verb|image_labels_train.csv|, with the following fields: Image ID (\verb|image_id|), radiologist ID (\verb|rad_ID|), and labels (\verb|labels|) for both the findings and diagnoses. Each image ID is associated with a vector of multiple labels corresponding to different pathologies, with positive pathologies encoded as ``\verb|1|'' and negative pathologies encoded as ``\verb|0|''. Similarly, the test set's bounding-box annotations and image-level labels were saved in the files \verb|annotations_test.csv| and \verb|image_labels_test.csv|, respectively.

\section*{Technical Validation}

The data de-identification process was controlled. Specifically, all DICOM meta-data was parsed and manually reviewed to ensure that all individually identifiable health information (PHI)~\cite{isola2019protected} of the children patients has been removed to meet the U.S. HIPAA~\cite{assistance2003summary} regulations. In addition, pixel values of all pediatric CXR scans were also carefully examined by human readers. During this review process, all scans were  manually reviewed case-by-case by a team of 10 human readers. A small number of images containing private textual information that had not been removed by our algorithm was excluded from the dataset. The manual review process also helped identify and discard out-of-distribution samples such as CXRs of adult patients, body parts other than the chest, low-quality images, or lateral CXRs that our machine learning classifier was not able to detect. A set of rules underlying our web-based annotation tool were developed to control the quality of the labeling process. These rules prevent human annotators from mechanical mistakes like forgetting to choose global labels or marking lesions on the image while choosing ``\verb|No finding|'' as the global label.

\section*{Usage Notes}

The PediCXR dataset was established for the purpose of developing and evaluating machine learning algorithms for detecting and localizing anomalies in pediatric CXR images. The dataset has been previously used in a study on the diagnosis of multiple diseases in pediatric patients~\cite{tran2021learning} and showed promising results. Specifically, the authors ~\cite{tran2021learning} introduced a deep learning network to detect common pulmonary pathologies on CXR
of pediatric patients. On the test set of 777 studies of the PediCXR dataset, the network yielded an area under the receiver operating characteristic (AUC) of 0.709 (95\% CI, 0.690–0.729). The sensitivity, specificity, and F1-score at the cutoff value are 0.722 (0.694–0.750), 0.579 (0.563–0.595), and 0.389 (0.373–0.405), respectively. However, they recognized that its performance remains low compared to medical experts. This work revealed the major challenge in learning disease features on pediatric CXR images using representation learning techniques, opening huge aspects for future research.\\ 
\\
The primary uses for which the PediCXR dataset was conceptualized include:

\begin{itemize}

  \item Developing and validating a predictive model for the classification of common thoracic diseases in pediatric patients.
  \item Developing and validating a predictive model for the localization of multiple abnormal findings on the pediatric chest X-ray scans.

\end{itemize}

\noindent Finally, the released dataset remains with limitations that still need to be addressed in the future, including:

\begin{itemize}
  \item The dataset did not contain clinical information associated with DICOM images, which is essential for the interpretation of CXR in children patients. 
  \item The number of examples for rare diseases (e.g., Congenital pulmonary airway malformation (CPAM), Congenital emphysema, Diagphramatic hernia, Mediastinal tumor, Pleuro-pneumonia, Situs inversus, Lung tumor) or findings (Emphysema, Edema, Calcification, Chest wall mass, Bronchectasis, Pleural thickening, Clavicle fracture, Pleuropulmonary mass, Paraveterbral mass, etc.) are limited. Hence, training supervised learning algorithms, which requires a large-scale annotated dataset, on the PediCXR dataset to diagnose the rare diseases and findings is not reliable. 
\end{itemize}

To download and use the PediCXR , users are required to accept the \href{https://physionet.org/content/mimic-cxr/view-license/2.0.0/}{PhysioNet Credentialed Health Data License 1.5.0}. By accepting this license, users agree that they will not share access to the dataset with anyone else. For any publication that explores this resource, the authors must cite this original paper and release their code and models. 

\section*{Code Availability}
\label{sec:code}
This study used the following open-source repositories to load and process DICOM scans: Python 3.7.0 (\href{https://www.python.org/}{https://www.python.org/}); Pydicom 1.2.0 (\href{https://pydicom.github.io/}{https://pydicom.github.io/}); OpenCV-Python 4.2.0.34 (\href{https://pypi.org/project/opencv-python/}{https://pypi.org/project/opencv-python/}); and Python hashlib (\href{https://docs.python.org/3/library/hashlib.html}{https://docs.python.org/3/library/hashlib.html}). The code for data de-identification was made publicly available at \href{https://github.com/vinbigdata-medical/vindr-cxr}{https://github.com/vinbigdata-medical/vindr-cxr}. The code to train CNN classifier for the out-of-distribution task was made publicly available at \href{https://github.com/vinbigdata-medical/DICOM-Imaging-Router}{https://github.com/vinbigdata-medical/DICOM-Imaging-Router}. The VinDr Lab is an open source software and can be found at \href{https://vindr.ai/vindr-lab}{https://vindr.ai/vindr-lab}.

\bibliography{sample}

\begin{thebibliography}{10}
\urlstyle{rm}
\expandafter\ifx\csname url\endcsname\relax
  \def\url#1{\texttt{#1}}\fi
\expandafter\ifx\csname urlprefix\endcsname\relax\def\urlprefix{URL }\fi
\expandafter\ifx\csname doiprefix\endcsname\relax\def\doiprefix{DOI: }\fi
\providecommand{\bibinfo}[2]{#2}
\providecommand{\eprint}[2][]{\url{#2}}

\bibitem{GBD2015}
\bibinfo{author}{Collaborators, G. .~L.}
\newblock \bibinfo{journal}{\bibinfo{title}{Estimates of the global, regional,
  and national morbidity, mortality, and aetiologies of lower respiratory tract
  infections in 195 countries: a systematic analysis for the global burden of
  disease study 2015}}.
\newblock {\emph{\JournalTitle{The Lancet Infectious Diseases}}}
  \textbf{\bibinfo{volume}{17}}, \bibinfo{pages}{1133--1161}
  (\bibinfo{year}{2017}).

\bibitem{unicef2006}
\bibinfo{author}{Wardlaw, T.~M.}, \bibinfo{author}{Johansson, E.~W.},
  \bibinfo{author}{Hodge, M.}, \bibinfo{author}{Organization, W.~H.} \&
  \bibinfo{author}{(UNICEF), U. N. C.~F.}
\newblock \bibinfo{title}{Pneumonia : the forgotten killer of children}
  (\bibinfo{year}{2006}).

\bibitem{hart2019pediatric}
\bibinfo{author}{Hart, A.} \& \bibinfo{author}{Lee, E.~Y.}
\newblock \bibinfo{journal}{\bibinfo{title}{Pediatric chest disorders:
  Practical imaging approach to diagnosis}}.
\newblock {\emph{\JournalTitle{Diseases of the Chest, Breast, Heart and Vessels
  2019-2022}}} \bibinfo{pages}{107--125} (\bibinfo{year}{2019}).

\bibitem{pcxr}
\bibinfo{title}{Chest radiograph (pediatric)}.
\newblock
  \bibinfo{howpublished}{\url{https://radiopaedia.org/articles/chest-radiograph-paediatric}}.
\newblock \bibinfo{note}{Accessed: 2021-09-24}.

\bibitem{du2002observer}
\bibinfo{author}{Du~Toit, G.}, \bibinfo{author}{Swingler, G.} \&
  \bibinfo{author}{Iloni, K.}
\newblock \bibinfo{journal}{\bibinfo{title}{Observer variation in detecting
  lymphadenopathy on chest radiography}}.
\newblock {\emph{\JournalTitle{International Journal of Tuberculosis and Lung
  Disease}}} \textbf{\bibinfo{volume}{6}}, \bibinfo{pages}{814--817}
  (\bibinfo{year}{2002}).

\bibitem{wang2017chestx}
\bibinfo{author}{Wang, X.} \emph{et~al.}
\newblock \bibinfo{title}{{ChestX-ray8: Hospital-scale chest X-ray database and
  benchmarks on weakly-supervised classification and localization of common
  thorax diseases}}.
\newblock In \emph{\bibinfo{booktitle}{Proceedings of the IEEE Conference on
  Computer Vision and Pattern Recognition (CVPR)}},
  \bibinfo{pages}{2097--2106}, \url{https://doi.org/10.1109/CVPR.2017.369}
  (\bibinfo{year}{2017}).

\bibitem{bustos2019padchest}
\bibinfo{author}{Bustos, A.}, \bibinfo{author}{Pertusa, A.},
  \bibinfo{author}{Salinas, J.-M.} \& \bibinfo{author}{de~la Iglesia-Vay{\'a},
  M.}
\newblock \bibinfo{journal}{\bibinfo{title}{{Padchest: A large chest X-ray
  image dataset with multi-label annotated reports}}}.
\newblock {\emph{\JournalTitle{arXiv preprint arXiv:1901.07441}}}
  (\bibinfo{year}{2019}).

\bibitem{irvin2019chexpert}
\bibinfo{author}{Irvin, J.} \emph{et~al.}
\newblock \bibinfo{title}{{CheXpert: A large chest radiograph dataset with
  uncertainty labels and expert comparison}}.
\newblock In \emph{\bibinfo{booktitle}{Proceedings of the AAAI Conference on
  Artificial Intelligence}}, vol.~\bibinfo{volume}{33},
  \bibinfo{pages}{590--597} (\bibinfo{year}{2019}).

\bibitem{johnson2019mimic}
\bibinfo{author}{Johnson, A.~E.} \emph{et~al.}
\newblock \bibinfo{journal}{\bibinfo{title}{{MIMIC-CXR, a de-identified
  publicly available database of chest radiographs with free-text reports}}}.
\newblock {\emph{\JournalTitle{Scientific Data}}} \textbf{\bibinfo{volume}{6}},
  \bibinfo{pages}{317}, \url{https://doi.org/10.1038/s41597-019-0322-0}
  (\bibinfo{year}{2019}).

\bibitem{nguyen2020vindr}
\bibinfo{author}{Nguyen, H.~Q.} \emph{et~al.}
\newblock \bibinfo{journal}{\bibinfo{title}{Vindr-cxr: An open dataset of chest
  x-rays with radiologist's annotations}}.
\newblock {\emph{\JournalTitle{arXiv preprint arXiv:2012.15029}}}
  (\bibinfo{year}{2020}).

\bibitem{jaeger2014two}
\bibinfo{author}{Jaeger, S.} \emph{et~al.}
\newblock \bibinfo{journal}{\bibinfo{title}{{Two public chest X-ray datasets
  for computer-aided screening of pulmonary diseases}}}.
\newblock {\emph{\JournalTitle{Quantitative Imaging in Medicine and Surgery}}}
  \textbf{\bibinfo{volume}{4}}, \bibinfo{pages}{475--477},
  \url{https://dx.doi.org/10.3978\%2Fj.issn.2223-4292.2014.11.20}
  (\bibinfo{year}{2014}).

\bibitem{tabik2020covidgr}
\bibinfo{author}{Tabik, S.} \emph{et~al.}
\newblock \bibinfo{journal}{\bibinfo{title}{Covidgr dataset and covid-sdnet
  methodology for predicting covid-19 based on chest x-ray images}}.
\newblock {\emph{\JournalTitle{IEEE journal of biomedical and health
  informatics}}} \textbf{\bibinfo{volume}{24}}, \bibinfo{pages}{3595--3605}
  (\bibinfo{year}{2020}).

\bibitem{rajpurkar2017chexnet}
\bibinfo{author}{Rajpurkar, P.} \emph{et~al.}
\newblock \bibinfo{journal}{\bibinfo{title}{{CheXNet: Radiologist-level
  pneumonia detection on chest X-rays with deep learning}}}.
\newblock {\emph{\JournalTitle{arXiv preprint arXiv:1711.05225}}}
  (\bibinfo{year}{2017}).

\bibitem{rajpurkar2018deep}
\bibinfo{author}{Rajpurkar, P.} \emph{et~al.}
\newblock \bibinfo{journal}{\bibinfo{title}{{Deep learning for chest radiograph
  diagnosis: A retrospective comparison of the CheXNeXt algorithm to practicing
  radiologists}}}.
\newblock {\emph{\JournalTitle{PLoS Medicine}}} \textbf{\bibinfo{volume}{15}},
  \bibinfo{pages}{e1002686}, \url{https://doi.org/10.1371/journal.pmed.1002686}
  (\bibinfo{year}{2018}).

\bibitem{majkowska2020chest}
\bibinfo{author}{Majkowska, A.} \emph{et~al.}
\newblock \bibinfo{journal}{\bibinfo{title}{{Chest radiograph interpretation
  with deep learning models: Assessment with radiologist-adjudicated reference
  standards and population-adjusted evaluation}}}.
\newblock {\emph{\JournalTitle{Radiology}}} \textbf{\bibinfo{volume}{294}},
  \bibinfo{pages}{421--431}, \url{https://doi.org/10.1148/radiol.2019191293}
  (\bibinfo{year}{2020}).

\bibitem{rajpurkar2020chexpedition}
\bibinfo{author}{Rajpurkar, P.} \emph{et~al.}
\newblock \bibinfo{journal}{\bibinfo{title}{{CheXpedition: Investigating
  generalization challenges for translation of chest X-ray algorithms to the
  clinical setting}}}.
\newblock {\emph{\JournalTitle{arXiv preprint arXiv:2002.11379}}}
  (\bibinfo{year}{2020}).

\bibitem{tang2020automated}
\bibinfo{author}{Tang, Y.-X.} \emph{et~al.}
\newblock \bibinfo{journal}{\bibinfo{title}{Automated abnormality
  classification of chest radiographs using deep convolutional neural
  networks}}.
\newblock {\emph{\JournalTitle{npj Digital Medicine}}}
  \textbf{\bibinfo{volume}{3}}, \bibinfo{pages}{1--8},
  \url{https://doi.org/10.1038/s41746-020-0273-z} (\bibinfo{year}{2020}).

\bibitem{pham2020interpreting}
\bibinfo{author}{Pham, H.~H.}, \bibinfo{author}{Le, T.~T.},
  \bibinfo{author}{Tran, D.~Q.}, \bibinfo{author}{Ngo, D.~T.} \&
  \bibinfo{author}{Nguyen, H.~Q.}
\newblock \bibinfo{journal}{\bibinfo{title}{{Interpreting chest X-rays via CNNs
  that exploit hierarchical disease dependencies and uncertainty labels}}}.
\newblock {\emph{\JournalTitle{arXiv preprint arXiv:1911.06475}}}
  (\bibinfo{year}{2020}).

\bibitem{KERMANY20181122}
\bibinfo{author}{Kermany, D.~S.} \emph{et~al.}
\newblock \bibinfo{journal}{\bibinfo{title}{Identifying medical diagnoses and
  treatable diseases by image-based deep learning}}.
\newblock {\emph{\JournalTitle{Cell}}} \textbf{\bibinfo{volume}{172}},
  \bibinfo{pages}{1122--1131.e9},
  \url{https://doi.org/10.1016/j.cell.2018.02.010} (\bibinfo{year}{2018}).

\bibitem{Chen2020}
\bibinfo{author}{Chen, K.-C.} \emph{et~al.}
\newblock \bibinfo{journal}{\bibinfo{title}{{Diagnosis of common pulmonary
  diseases in children by X-ray images and deep learning}}}.
\newblock {\emph{\JournalTitle{Scientific Reports}}}
  \textbf{\bibinfo{volume}{10}}, \bibinfo{pages}{1--9} (\bibinfo{year}{2020}).

\bibitem{gordon2019explainable}
\bibinfo{author}{Gordon, L.}, \bibinfo{author}{Grantcharov, T.} \&
  \bibinfo{author}{Rudzicz, F.}
\newblock \bibinfo{journal}{\bibinfo{title}{Explainable artificial intelligence
  for safe intraoperative decision support}}.
\newblock {\emph{\JournalTitle{JAMA surgery}}} \textbf{\bibinfo{volume}{154}},
  \bibinfo{pages}{1064--1065} (\bibinfo{year}{2019}).

\bibitem{assistance2003summary}
\bibinfo{author}{{US Department of Health and Human Services}}.
\newblock \bibinfo{title}{{Summary of the HIPAA privacy rule}}.
\newblock
  \bibinfo{howpublished}{\url{https://www.hhs.gov/hipaa/for-professionals/privacy/laws-regulations/index.html}}
  (\bibinfo{year}{2003}).

\bibitem{isola2019protected}
\bibinfo{author}{Isola, S.} \& \bibinfo{author}{Al~Khalili, Y.}
\newblock \bibinfo{title}{{Protected Health Information (PHI)}}.
\newblock
  \bibinfo{howpublished}{\url{https://www.ncbi.nlm.nih.gov/books/NBK553131/}}
  (\bibinfo{year}{2019}).

\bibitem{pham2021dicom}
\bibinfo{author}{Pham, H.~H.}, \bibinfo{author}{Do, D.~V.} \&
  \bibinfo{author}{Nguyen, H.~Q.}
\newblock \bibinfo{journal}{\bibinfo{title}{Dicom imaging router: An open deep
  learning framework for classification of body parts from dicom x-ray scans}}.
\newblock {\emph{\JournalTitle{arXiv preprint arXiv:2108.06490}}}
  (\bibinfo{year}{2021}).

\bibitem{nguyen2021vindr}
\bibinfo{author}{Nguyen, N.~T.} \emph{et~al.}
\newblock \bibinfo{journal}{\bibinfo{title}{Vindr lab: A data platform for
  medical ai}}.
\newblock {\emph{\JournalTitle{URL: https://github.
  com/vinbigdata-medical/vindr-lab}}}  (\bibinfo{year}{2021}).

\bibitem{tran2021learning}
\bibinfo{author}{Tran, T.~T.} \emph{et~al.}
\newblock \bibinfo{title}{Learning to automatically diagnose multiple diseases
  in pediatric chest radiographs using deep convolutional neural networks}.
\newblock In \emph{\bibinfo{booktitle}{IEEE Conference on Computer Vision and
  Pattern Recognition Workshop (ICCV 2021)}} (\bibinfo{year}{2021}).

\end{thebibliography}

\section*{Acknowledgements}

The collection of this dataset was funded by the Smart Health Center, VinBigData JSC. The authors would like to acknowledge the Phu Tho Obstetric \& Pediatric Hospital for agreeing to make the PediCXR dataset publicly available. We are especially thankful to Anh T. Nguyen, Huong T.T. Nguyen, Ngan T.T. Nguyen for their helps in the data collection and labeling process. 

\section*{Author contributions}

H.Q.N. and H.H.P designed the study; T.T.T. performed the data de-identification; H.Q.N., and H.H.P. wrote the paper; all authors reviewed the manuscript.

\section*{Competing interests} 

This work was funded by the Vingroup JSC. The funder had no role in study design, data collection and analysis, decision to publish, or preparation of the manuscript.

\end{document}